# IR color separation in transmission through gratings on (110) silicon: FTIR experiment versus theory

**Mark Auslender and Shlomo Hava**
Ben-Gurion University of the Negev, Department of Electrical and Computer
Engineering, POB 653, Beer-Sheva 84105, Israel
marka@eesrv.bgu.ac.il,  hava@eesrv.bgu.ac.il

When supporting several orders, diffraction gratings demonstrate the phenomenon of spectral filtering in zero-order beam that was studied in a narrow visible range for glass gratings.[1] The reason for such a behavior is an exchange of the radiation power between the zero- and the higher-order beams. For special grating dimensions the zero-order beam intensity may almost vanish at some wavelengths and may be very small in their vicinity that creates dark bands (color separation). We have studied this phenomenon in a wide mid-infrared range for gratings micromachined on the top of (110) silicon wafer using two different measurement techniques and numerical simulations.[2-4] We started with an FTIR study,[2] and proceeded using a monochromator-based setup.[3,4] However, four samples in those studies did not demonstrate the best possible performance. In this summary we report on the study of the phenomenon with much wider set of grating samples (14 ones including the above four), most of which show excellent color separation. Due to its advantages, FTIR spectrometry has been the method of choice in the present study.

The grating samples were fabricated using an updated technology based on anisotropic-wet-etching micromachining of crystalline silicon in alkaline solutions.[5] The process exploited cleaned and thermally oxidized two-side polished 500μm thick (110) lightly doped n – Si wafer, and a mask containing 14 square patterns of area of 1×1 $cm^2$ each, filled with grating lines. By the use of photolithography and oxide etching the patterns of thin grating in oxide were produced. After photoresist removal the wafer was cut to obtain the samples. The cuts were etched in 40% KOH aqueous solution at the temperature of 40°C (oxide grating itself served as the mask) that produced the lamellae grooves with vertical walls and flat bottom. Different time of etching controlled different groove depths of the samples.

Using a computerized Perkin Elmer 1600 FTIR spectrometer with the wavenumber resolution of 2 $cm^{-1}$ the non-polarized normal-incidence transmission spectra of the grating samples and a smooth-surface sample cut from the master wafer were measured in the wavelength interval 2.2μm $\leq \lambda \leq$ 22μm. Measured samples were coated by a black paper with two coincident circular windows on both sides and placed in standard Perkin Elmer mounts. The intensity of radiation transmitted through the windows in empty cover was used as the background signal. To spare time 32 scans were used.

Measured spectrum of the smooth-surface sample is in excellent agreement with that reported in the literature. Its peculiar feature is two pronounced dip-like bands at about 1107 $cm^{-1}$ and 610 $cm^{-1}$. The bands are attributed to two dominant infrared absorption mechanisms in bulk silicon: due to transitions in the complex Si–$O_2$ and the lattice phonons, respectively.[6] A typical feature of the grating samples' spectra (the examples are shown in Figs.1 and 2) is quasi-periodic transmittance variation versus wavenumber at $\lambda < \Lambda$ ($\Lambda$ is the grating period). The period of the oscillations depends mainly on the grating groove depth $H$. Depending on $\Lambda/W$ ($W$ is the grating groove width) and $H$, the transmittance may be very low (ideally zero) at the minima. The repetition of the dark and bright bands creates the color separation (see figures).

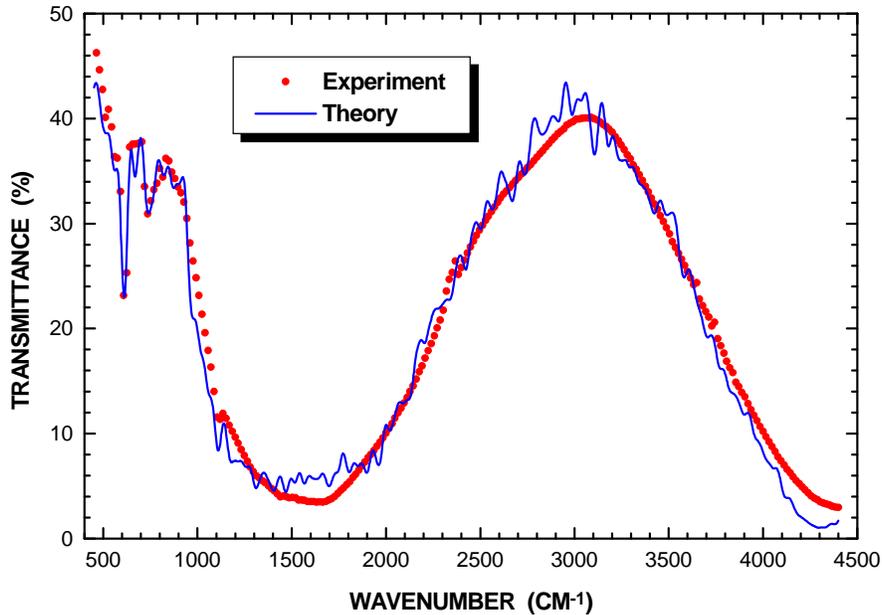

Fig. 1: Measured and calculated transmittance of grating sample with $\Lambda = 10.6\,\mu m$, $W = 5.5\,\mu m$ and etch-time estimated $H_e = 1.7\,\mu m$. Numerical simulation corrects the groove depth to $H_s = 1.4\,\mu m$

In the cases where the inherent absorption dip position is embedded in the multi-order region ($\lambda < \Lambda$) it is hardly seen on the grating spectrum. This concerns the $Si - O_2$ - related dip in Fig.1 and the both dips in Fig.2. However, if the dip position lies in zero-order region the ($\lambda > \Lambda$) the transmission in this band and nearby may be much lower than that for the smooth-surface sample, as happens with the phonon-related dip in Fig.1. The reason for this phenomenon is the absorption by the non-specular orders in the substrate, which is higher than that of the zero order due to the longer optical path.

The theoretical account for the observed spectra consists of two principal parts. The first is numerical simulations using software we developed on the base of S-matrix propagation algorithm implemented with an updated Fourier-series expansion procedure.[7] The program calculates the polarized (transmitted and reflected) diffraction efficiencies of a multilayer grating structure, stable and convergent regardless of how many grating and homogeneous layers the structure consists of, and how thick each layer is, and how many Fourier space harmonics are retained in the simulation. The non-polarized diffraction efficiencies are obtained by taking half the sum of p-polarized and s-polarized ones. The convergence is vitally important since potentially many diffracted orders may propagate. For example, for $\Lambda = 22\,\mu m$ (Fig.2) at $\lambda = 2.2\,\mu m$ there exist 21 orders in the air and 69 orders and in the substrate. The calculated transmittance contains dense interference fringes, whereas the experimental data are almost fringe-free. Consequently, the second part is a procedure of smoothing the fringes we developed using a signal-processing approach based on low-pass digital filtering.

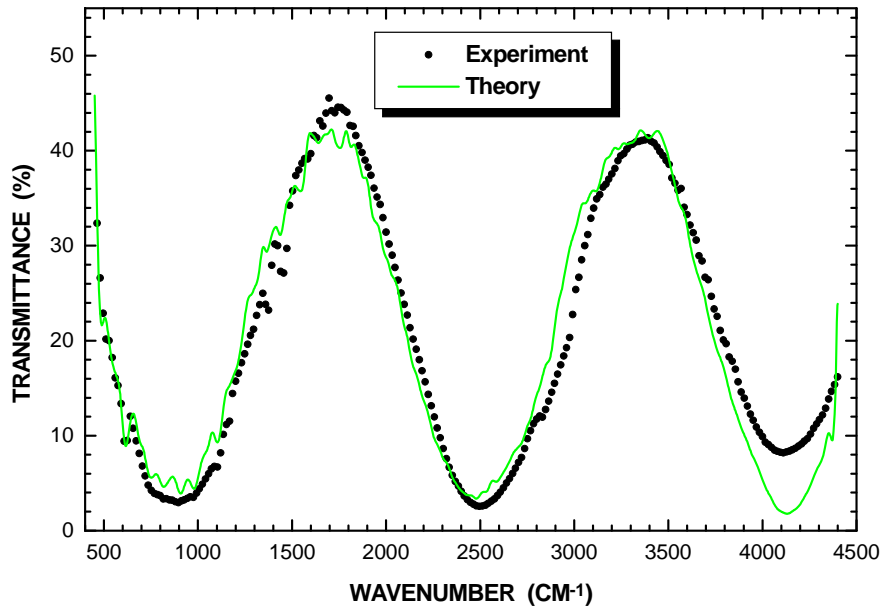

Fig. 2: Measured and calculated transmittance of grating sample with $\Lambda = 22\mu m$, $W = 9\ \mu m$ and etch-time estimated $H_e = 2.8\ \mu m$. Numerical simulation corrects the groove depth to $H_s = 2.5\ \mu m$

We have revealed by comparison that the simulation results fit the data provided that the grating groove depth in the most samples is corrected. In many cases the correction is not drastic (see captions to figures), but in some cases $H$'s estimated from etching time turn out to be $2 \div 3\mu m$ more than needed to fit the experimental data. This under-etching may be due to slowing the etch rate or spoiling the etching solution. After the etch depth correction the theory and the experiment are in pretty good agreement as seen from Figs.1 and 2. Thus, the combination of measurement and numerical simulation appears to be a cheap tool for checking the etch depth values in grating production.